\documentclass[11pt, a4paper]{article}
\usepackage{geometry} 
\usepackage{courier}
\usepackage[bottom]{footmisc}
\usepackage{apacite}
\usepackage{natbib}
\usepackage{amsmath}
\usepackage{amssymb}
\usepackage{amsthm}
\usepackage{graphicx}
\usepackage[hidelinks]{hyperref}
\usepackage{titlesec}
\usepackage{pifont}
\usepackage{comment}
\usepackage{setspace}
\usepackage{enumerate}
\usepackage{bm}
\usepackage{subcaption}
\usepackage{float}
\usepackage{booktabs}
\usepackage{threeparttable}
\usepackage[linesnumbered, lined, boxed, ruled, vlined]{algorithm2e}

\titlelabel{\thetitle.\,\,}
\newcommand{\citew}[1]{\citeauthor{#1}'s (\citeyear{#1})}
\newtheorem{theorem}{Theorem}
\newtheorem{lemma}{Lemma}
\newtheorem{definition}{Definition}
\newtheorem{remark}{Remark}
\newtheorem{assumption}{Assumption}
\newtheorem{assumptionalt}{Assumption}[assumption]

\newenvironment{assumptionp}[1]{
  \renewcommand\theassumptionalt{#1}
  \assumptionalt
}{\endassumptionalt}

\parskip=\baselineskip
\title{Generating Synthetic Data with Locally Estimated Distributions for Disclosure Control\thanks{This study is part of \citew{kalay2024thesis} thesis entitled ``Essays on Administrative Data Methodologies'' at the University of Queensland, School of Economics. The implementation of the proposed algorithms are available as at The Python Package Index: \href{https://pypi.org/project/synloc/}{pypi.org/project/synloc/}.}}
\author{Ali Furkan Kalay\thanks{\textit{alifurkan.kalay@mq.edu.au}, Macquarie University Centre for Health Economy}}

\date{\today }

\begin{document}

\maketitle

\begin{abstract}
  Sensitive datasets are often underutilized in research and industry due to privacy concerns, limiting the potential of valuable data-driven insights. Synthetic data generation presents a promising solution to address this challenge by balancing privacy protection with data utility. This paper introduces a new approach to mitigate privacy risks associated with outlier observations in synthetic datasets: the Local Resampler (LR). The LR leverages the $k$-nearest neighbors algorithm to generate synthetic data while minimizing disclosure risks by underrepresenting outliers, even when they are not detectable in marginal distributions. Theoretical and empirical analyses demonstrate that the LR effectively mitigates outlier-driven disclosure risks, and accurately replicates multimodal, skewed, and non-convex support distributions. The semiparametric nature of the LR ensures a low computational burden and works efficiently even with small samples. By parameterizing the balance between privacy risks and data utility, this approach promotes broader access to sensitive datasets for research.
\end{abstract}

\newpage

\section{Introduction} \label{sec:introduction}

Sensitive datasets, such as financial records, healthcare data, and administrative data, hold significant potential for generating data-driven insights. Yet, stringent privacy regulations and the need to protect sensitive information often limits access to these resources. Synthetic data generation methods offer an alternative, allowing researchers to work with stochastic-model-generated datasets that preserve statistical characteristics while minimizing privacy risks. The optimal synthetic data methods must meet two critical standards: (1) accurate representation of the original dataset's statistical properties, and (2) robust disclosure risk control to protect the identity of individuals, especially those who exhibit atypical patterns (outliers). While synthetic data generation techniques are evolving rapidly, inherent control of disclosure risks remains a challenge. The lack of robust statistical disclosure control methods causes authorities to adopt cautious approaches when developing synthetic data infrastructures, limiting the utility of sensitive datasets.

This paper introduces a new approach, the Local Resampler (LR), which leverages the $k$-nearest neighbor algorithm for the generation of synthetic datasets.  LR demonstrates two key advantages: (1) the ability to accurately replicate complex distributions (e.g., multimodal, skewed and non-convex support distributions) with minimal hyperparameter tuning, and (2) the capacity to minimize disclosure risks by underrepresenting outliers in the synthetic data.  In contrast to benchmark approaches, the latter feature eliminates the need for analysts\footnote{I refer to the ``analyst'' as the individual responsible for creating synthetic data to ensure disclosure control, while maintaining the utility of the dataset. This means that any statistical analysis performed on the synthetic data should closely resemble the results obtained from the original sample.} to resort to time-consuming manual trimming processes often limited to marginal distributions, which are insufficient for comprehensive outlier identification.

While the importance of outliers in synthetic data literature is recognized \citep{sweeney2002k, drechsler2011textbook,Bowen2021Protecting,jordon2022synthetic}, existing methods lack the ability to statistically control the disclosure risks they pose. This paper uniquely addresses risks driven specifically by outliers within joint distributions and show that LR inherently handles such outlier-driven disclosure risk. 

The concept of synthetic data in research has been discussed and explored for decades, initiated by \citet{rubin1993statistical}. A notable example is The Longitudinal Business Database (LBD). The synthetic LBD allows researchers to obtain preliminary results and test compatibility of their code before submitting scripts (STATA and SAS) to the United States Census Bureau. This paper specifically addresses the risks posed by outliers within joint distributions and shows that LR inherently handles such outlier-driven disclosure risks.

The use of sensitive datasets for research purposes is becoming increasingly common. Consequently, there will be a growing demand for infrastructure similar to the synthetic Longitudinal LBD and, hence, for reliable synthetic data methods. For instance, a search for ``administrative data'' on \textit{sciencedirect.com} yielded 3,206 articles in the year 2000. This number grew to 26,307 results in 2023.\footnote{Indeed, this surge is mainly driven by the overall increase in publications of academic papers. Figure \ref{fig:academic_publications} isolates this trend and displays the ratio of ``administrative data'' search results on \textit{sciencedirect.com} to total search outcomes by year. The proportions in 2023 are threefold higher than in 2000. Not surprisingly, this increase is especially pronounced in data-oriented disciplines, such as economics, in comparison to other fields across all years. This likely reflects the particular value of administrative data for policy-relevant research.} Research in health is another key domain often using sensitive datasets (e.g., \citet{chen2021synthetic, hernandez2022synthetic}). These datasets must be protected due to privacy concerns; however, they also hold significant potential for providing valuable insights. Reliable methods for disclosure control are essential for ensuring robust data governance practices that balance research potential with the ethical and responsible use of sensitive information.

LR is a generic approach aiming to address this demand by allowing users to control disclosure risks for data privacy. I illustrate (Appendix \ref{app:clustering}) how the locality of the distributions can be defined using other methods such as clustering algorithms. The LR algorithm is especially suitable for statistical disclosure control, as it has a unique property of being biased towards the sample mean (or cluster means in multimodal distributions), which inherently reduces disclosure risk by mitigating the impact of outliers that typically require manual intervention. This property is particularly important as it automates the process of safeguarding data privacy. Lastly, LR does not require large samples to work efficiently, unlike the deep learning algorithms, which are computationally burdensome and difficult to optimize.

Section \ref{sec:literature} reviews existing methods for creating synthetic data and highlights their broader applications. Section \ref{methodology} details the LR algorithm, along with its assumptions, limitations, and possible extensions. Section \ref{sec:simulations} demonstrates the application of LR using artificial datasets. Finally, Section \ref{sec:conclusion} concludes the paper and suggests directions for future research and applications.

\section{Literature Review}\label{sec:literature}

\citet{rubin1993statistical} proposed a multiple imputation approach to create synthetic versions of the microdata to ``honor the confidentiality constraints.'' Since then, the multiple imputation approach has been further developed.\footnote{See \citet{raghunathan2021synthetic} for a literature review.} Various software packages are available to create synthetic data today, such as \textbf{synthpop} \citep{nowok2016synthpop} and \textbf{SDV} \citep{SDVpackage}. \textbf{synthpop} adapts the multiple imputation approach suggested by \citet{rubin1993statistical, raghunathan2003multiple}, and allows incorporating both parametric and nonparametric models.\footnote{The package offers considerable flexibility to users, such as handling and synthesizing missing values, selecting synthesis sequences, stratifying the sample, sampling from predictive posterior distributions of the models, and so on. The theoretical aspects of \textbf{synthpop} package were discussed in \citet{raghunathan2003multiple, raab2014simplified}.} \textbf{SDV}, on the other hand, includes various methods, such as copula and deep learning algorithms.

Multiple imputation methods replace observed values with model-predicted values as if they were missing. These methods are well-established for addressing missing data problems \citep{van1999flexible, stekhoven2012missforest}. In the context of synthetic data generation, the approach differs slightly: the Analyst can choose separate models for each variable, and nonparametric methods are often preferred over parametric models to capture the nonlinearity and complexity of the original data distribution \citep{raab2014simplified, drechsler2011empirical}. However, nonparametric methods potentially overfit the data and outliers can be a particular privacy risk in such circumstances. Therefore, analysts are encouraged to trim outliers from the marginal distributions before generating synthetic data. This practice, while being efficient, is not sufficient to control such risks as shown in Section \ref{sec:trimming}.

An alternative approach involves estimating a joint distribution and drawing values from it, rather than sequentially estimating posterior predictive distributions for each variable. Copulas are a preferred solution due to their flexibility, accommodating both parametric (e.g., Gaussian copula) and nonparametric distributions (e.g., Vine copula) \citep{sun2019learning}. However, copula-based models can be challenging to generalize, requiring appropriate specification for marginal distributions and struggling to synthesize continuous and discrete variables simultaneously.

Deep learning algorithms, such as Generative Adversarial Networks (GAN) \citep{goodfellow2013generative} and Variational Autoencoders (VAE) \citep{kingma2013auto}, have also been employed for synthetic data generation. While these methods can model complex data distributions, they are computationally intensive and difficult to optimize. Recent advancements, such as the use of normalizing flows to synthesize tabular data \citep{kamthe2021copula}, have improved their applicability. These methods can synthesize mixed data types and replicate distributions with non-convex support.

Distance-based methods like $k$-Nearest Neighbors ($k$NN) have been utilized in various ways for creating synthetic data. \citet{chawla2002smote} introduced the Synthetic Minority Over-sampling Technique (SMOTE), which generates synthetic samples by interpolating between neighboring observations. This technique is primarily used to augment imbalanced datasets in classification problems. Numerous extensions and variants of SMOTE have been developed for different purposes; see \citet{smotecomparison} for a comprehensive overview.

Recently, \citet{sivakumar2022synthetic} proposed a method combining Mega-Trend Diffusion \citep{li2007using} and $k$NN, named \emph{k}-Nearest Neighbor Mega-Trend Diffusion. One of the main advantages of this approach is its ability to create high-quality synthetic datasets from small samples. It resembles the Local Resampling (LR) approach with $k$NN, but differs significantly in how $k$NN is employed.

The use of clustering algorithms like K-means for synthetic data generation is not a new concept. The LR approach that incorporates K-means shares similarities with mixture models, which have been previously applied for synthetic data generation \citep{chokwitthaya2020applying}. However, estimating mixture models can be computationally demanding. A related technique is the K-means SMOTE algorithm introduced by \citet{douzas2018improving}, which utilizes K-means to enhance imbalanced datasets for classification purposes, similar to other SMOTE variants.

The multiple imputation method, as originally proposed by \citet{rubin1993statistical}, offers a straightforward implementation and demonstrates robustness compared to many alternative methods. Nonparametric approaches within multiple imputation are capable of effectively capturing asymmetric and nonlinear distributions for both discrete and continuous data \citep{drechsler2011empirical}. Moreover, \citet{raab2014simplified} provided a theoretical foundation for making valid inferences from synthetic samples—an aspect often lacking in other approaches. Given these benefits, \textbf{synthpop} is used as the primary benchmark method in this study, representing this approach. Additionally, methods available in the \textbf{SDV} package \citep{SDVpackage}, which is continuously evolving and includes diverse techniques such as copulas and deep learning, are also evaluated. \textbf{synthpop} and \textbf{SDV} were selected due to their popularity and their focus on generating synthetic data for disclosure control, which aligns well with the objectives of this paper.

While $k$NN and other distance-based methods are widely recognized for their potential in synthetic data generation—exemplified by techniques such as SMOTE and $k$NN diffusion models—this study reconceptualizes the SMOTE algorithm specifically to address synthetic data generation with a focus on privacy preservation. The LR approach is essentially a variant of the SMOTE algorithm, but it has been adapted to mitigate disclosure risks posed by outliers, particularly within joint distributions—an area largely overlooked in existing literature. This study conceptualizes the outlier problem in synthetic datasets and demonstrates the effectiveness of LR both theoretically and empirically. By establishing the theoretical and practical advantages of using $k$NN within the LR framework, this study illustrates how LR can be parameterized to balance data utility (i.e., replication accuracy) with privacy risks, all while maintaining a low computational cost.

\section{Local Resampler Algorithm}\label{methodology}

Let $\bm{x}_i$ be a $p$-dimensional vector, where $i$ represents the observation in our sample $S$ of size $n$. Our goal is to create a synthetic sample $\{\hat{\bm{x}}_i\}_{i=1}^{n'}$ of size $n'$ that has similar distributional properties to $S$. A more straightforward approach would involve estimating the distribution, $\hat{F}(\cdot)$, from sample $S$ and then resampling $\{\hat{\bm{x}}_i\}_{i=1}^{n'}$ from this estimated distribution where 

\begin{equation}
  \hat{\bm{x}}_i \sim \hat{F}(\cdot).
\end{equation}

Indeed, this approach can produce distributional properties similar to \( \bm{x}_i \) in some respects. However, a significant drawback is its strong dependence on distributional assumptions. For instance, copulas can be employed to estimate distributions \( F(\cdot) \), providing flexible choices for the marginal distribution of each variable in the dataset. Yet, this method necessitates a detailed analysis of the original sample, preprocessing of the variables, and the selection of the appropriate distribution family for each variable's marginal distribution.

Another challenge is managing high dimensionality. This can be addressed by relying on parametric distributions or employing dimension-reduction methods like functional Principal Component Analysis \citep{Meyer_Nagler_2021}. While existing approaches can benefit from the methodology presented in this paper, our findings demonstrate that LR performs well even under strong parametric distributional assumptions.

Let $d(\bm a, \bm  b)$ be the normalized Euclidean distance function measuring the distance between vectors $\bm a$ and $\bm b$; and $i(m)$ be the $m^{th}$ closest observation to $i$ such that,

\begin{equation}
  \sum _{j\in S} I\left \{  d(\bm {x} _i, \bm {x} _{i(m)} ) \ge d(\bm {x} _i, \bm {x} _{j} ) \right \} = m, \label{eq:m-closest}
\end{equation}

where $I(.)$ represents an indicator function, which takes the value of one if the given statement is true, and zero if it is not. Using the neighbors, we define the subsamples for each observation $i\in S$:

\begin{equation}
  S_i = \left \{ \bm { x} _{i(m)}\right \}_{m=1} ^{k},
\end{equation}

where $S_i$ is the set of neighbors of observation $i$. Note that $\bm{x}_{i(1)} = \bm x_i$, meaning that the first neighbor of each observation $i$ is itself. This approach aims to avoid information loss and ensure that all observations in $S$ are included in the subsamples.\footnote{Omitting $i$ from its neighboring set can enhance data privacy by reducing disclosure risks. This action will lead to further under-representation or even the absence of outliers in the set of subsamples \eqref{eq:set_subsample}. As a result, the synthetic sample will be further biased towards the sample mean.} The hyperparameter $k$ must be an integer greater than 2. The next sections will examine the impact of hyperparameter selection on privacy concerns and the efficiency of the synthetic data.

The set of subsamples is defined as follows:

\begin{equation}
  \mathbb{S} = \{ S_1, S_2, ..., S_n \}.\label{eq:set_subsample}
\end{equation}

After defining the subsamples, the distributional parameters are estimated for each subsample. Let $\mathbb{F}$ be the set of the estimated distributions for each $S_i$,

\begin{equation}
  \mathbb{F} = \left \{ \hat F_{S_{1}}, \hat F_{S_{2}}, ..., \hat F_{S_{n}}    \right \},\label{eq:set_of_dist}
\end{equation}

where $\hat F_{S_{i}}$ is the estimated distribution from subsample $S_{i}$. In order to create a synthetic sample with the required sample size, the set of the estimated distributions are resampled with replacement and equal probability,

\begin{equation}
  \mathbb{F'} = \left \{ \hat F_{S_{(1)}}, \hat F_{S_{(2)}}, ..., \hat F_{S_{(n)}}    \right \}.\label{eq:set_of_dist_new}
\end{equation}

This step disturbs the original sample and allows choosing $n' \not = n$. The analyst can opt not to resample the set of subsamples (i.e., $\mathbb{F}' = \mathbb{F}$). In this case, the synthetic sample will have the same size as the original sample. The last step is to draw observations from $\mathbb{F'}$ and create a synthetic sample with the required size $n'$,

\begin{equation}
  \left \{ \tilde{\bm x} _i\right \}_{i=1} ^{n'} = \left \{ \tilde{\bm x _{1}}, \tilde{\bm x _{2}}, ..., \tilde{\bm x _{n'}}\right \} \sim \mathbb{F} .\label{eq:draw_Sample}
\end{equation}

Algorithm \ref{alg:NN} summarizes the steps of the proposed methodology in this section. Henceforth, this algorithm shall be referred to as the $k$NN local resampler ($k$NN-LR). Appendix \ref{app:clustering} provides an alternative approach using K-means to define subsamples. 

\begin{algorithm}[!ht]
  \KwData{$\bm x$, of dimension $n\times p$ where $n$ is the sample size and $p$ is the number of variables.\tcp*{$\bm x_i$ is the $i^{th}$ observation from data}}
  \KwIn{$k$: the number of neighbors; $n'$: the required synthetic sample size.}
  \KwOut{$\tilde{\bm x}\leftarrow n'\times p$\tcp*{A synthetic dataset}}
  $d(i,j)\leftarrow $ Normalized Euclidean distance between $\bm{x} _i$ and $\bm{x} _j$ \;
  \tcc{Finding nearest neighbors}
  \For{$i\in \{1,2,...,n\}$}{
    \For{$m=1$ to $k$}{
      \tcc{Find $m^{th}$ closest observation to $i$}
      $i(m) \leftarrow$ $m^{th}$ closest observation to $i$ such that $m = \sum _{j=1}^n \mathbf{1}\left (  d(i,i(m)) \ge d(i,j) \right )$\;
    }
    \tcc{Set of neighbors for observation $i$:}
    $S_i \leftarrow \{i(1), i(2),...,i(k) \}$\;
  }
  \tcc{Estimating distributions for each subsample}
  \For{$i\in \{1,2,...,n\}$}{
    $\hat F_{S_i} \leftarrow $ Estimated distribution from subsample $S_i$\;
  }
  \tcc{Draw synthetic values from the estimated distributions with required size $n'$}
  \For{$i\in \{1,2,...,n'\}$}{
    $\mathbf{F}_i\leftarrow $ Randomly draw an estimated distribution from $\{\hat F_{S_1}, \hat F_{S_2},..., \hat F_{S_n}\}$\;
    $\tilde{\mathbf{x} _{i}}\leftarrow $ Draw a synthetic value from $\mathbf{F}_i$\;
  }
  \KwRet{$\tilde{\bm{x}} \leftarrow n'\times p$}
  \caption{Creating Synthetic Data with Nearest Neighbor Algorithm}
  \label{alg:NN}
\end{algorithm}

\subsection{Intuition of Local Resampling}\label{sec:intuition}

Consider the simplest scenario where the analyst wants to synthesize a sample with one variable. The distribution of this variable can be visualized with a histogram, which is, in fact, a semi-parametric approximation to a true probability density function. The values are stratified into bins, and the frequency of observations is then used to estimate the density function. For simplicity, let's assume that $x_i \in [0,1]$ drawn from an unknown distribution with a probability density function, $p(x)$. The analyst creates $M$ bins with equal-length partitions as follows,

\begin{equation}
B_1 = \left [0,\dfrac{1}{M} \right ), B_2 = \left [\dfrac{1}{M},\dfrac{2}{M} \right ),..., B_M = \left [\dfrac{M-1}{M},1\right ].
\end{equation}

The estimated density function for each bin is,

\begin{equation}
  \hat p_m(x) = \dfrac{M}{n}\sum_{i=1}^n I(x_i \in B_m), \quad m=1,2,...,M.
\end{equation}

\begin{figure}[!ht]
  \centering
  \includegraphics[width=\textwidth]{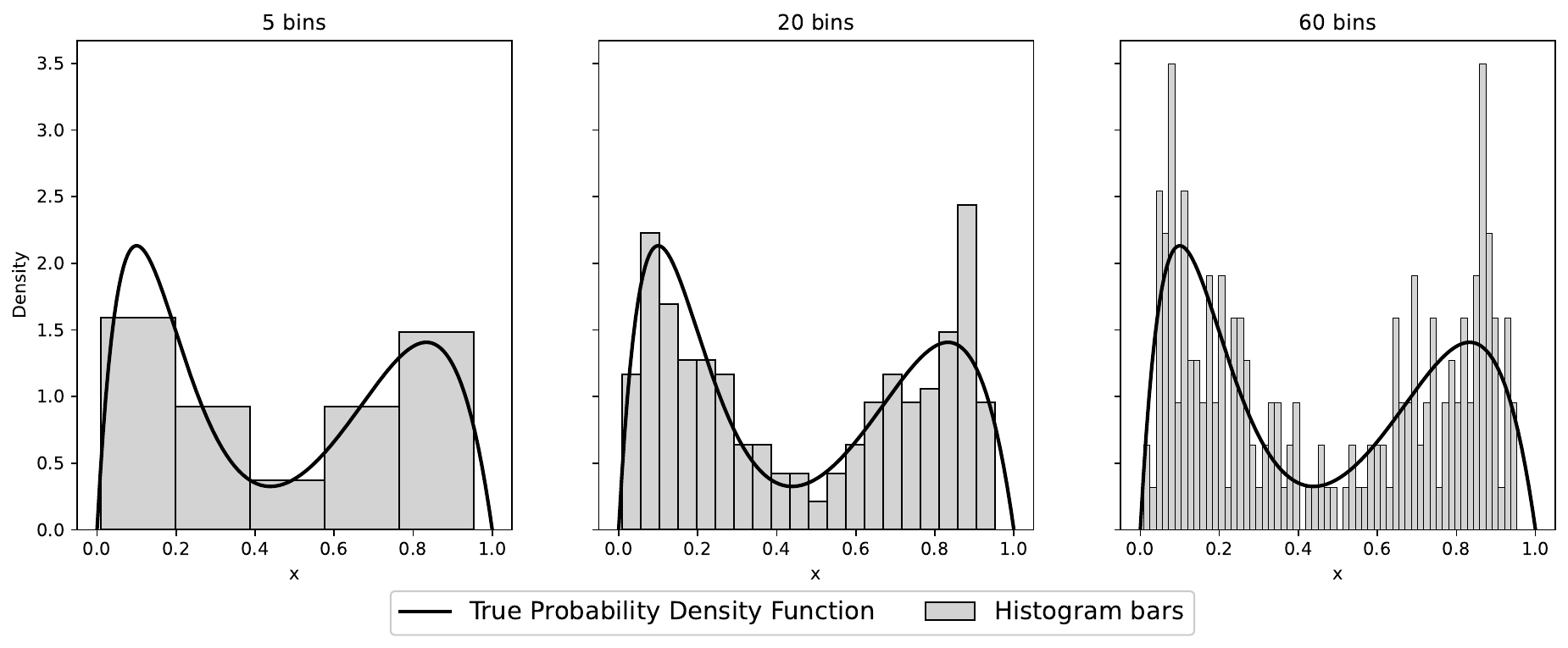}
  \caption{Histogram Approximations to a Distribution with Different Number of Bins}
  \label{fig:histograms}
\end{figure}

The selection of $M$ is a classical example of bias-variance trade-off. A large value of $M$ will lead to smaller bias, i.e., smaller $E[\hat p_m(x) - p(x)]$ for all $m\in \{1,2,...,M\}$; but will lead to larger variance, as shown in Figure \ref{fig:histograms}. 

Similarly, the proposed methodology, LR, partitions datasets by defining subsamples, then estimates parametric distributions for each subsample and draws \textit{synthetic} values from the estimated distributions. As with histogram approximation, bias will be reduced due to stratification, and dimensionality will be less of a concern due to parametric distributional assumptions.

Assume that we estimate the mean and standard deviation of the sample shown in Figure \ref{fig:histograms}, and then draw values from the estimated multivariate normal distribution using the estimated covariance matrix. The resulting synthetic data will not be bimodal like the original sample because of the imposed normality assumption. However, the histogram approximation can capture the two peaks in the data due to the partitioning of the sample space. The distribution within each partition is not necessarily as complex as the distribution of the entire sample. In fact, in this example, the bimodality of the distribution can be captured even with a flat probability density assumption (i.e., a uniform distribution).

The efficiency of the LR algorithm is driven by a similar intuition. We define subsamples based on their proximity to each other. Each subsample corresponds to a different sample space.\footnote{In contrast to histograms, which have distinct and non-overlapping bins, the subsample spaces in LR algorithm can overlap with each other.} The distributions of these subsamples are not necessarily as complex as the distribution of the entire sample. For example, the distribution of the sample could have a non-convex support but each subsample could have a convex support. Consequently, LR can replicate complex distributions with parametric distributional assumptions. Appendix \ref{sec:formal-assumptions} provides a formal description of the assumptions and discusses the implications of these assumptions. 

\subsection{A Special Property of Local Resampler with kNN: Trimming Outliers}\label{sec:trimming}

Outliers present a significant privacy risk because they are more identifiable in a synthetic sample compared to central (non-outlier) observations. \citet{sweeney2002k} proposed the concept of $k$-Anonymity to address privacy risks by ensuring that each individual is indistinguishable from at least $k$ others within a sample. However, achieving $k$-Anonymity is challenging when outliers are present, as they inherently differ from the majority of observations and are therefore significant privacy risks.

For example, consider a dataset of firms where one firm reports exceptionally high profits. If a nonparametric method like random forest is used to generate synthetic data, a common approach suggested by \citet{drechsler2011empirical}, the model may overfit the original sample. This overfitting can lead to synthetic observations that closely mimic the characteristics of the high-profit firm, potentially revealing sensitive information. To mitigate this risk, analysts might trim such extreme values from the marginal distributions, reducing the influence of outliers in the synthetic data generation process.

However, simply censoring or trimming outliers based on marginal distributions may not be sufficient. Consider a scenario in the same dataset where there is only one firm with both high profit and high debt values. Although neither profit nor debt values are distinguishable from those of other firms when viewed individually in the marginal distributions, the combination of high profit and high debt could distingush from other firms. This makes the firm an outlier in the joint distribution space. Identifying and handling such joint outliers remains a challenging task, as marginal distributions alone fail to capture the complexity of interactions between variables, and visualizing joint distributions is not feasible when there are more than three variables. I will illustrate such a scenario in Section \ref{sec:simulations}. 

$k$NN has a special property minimizing such risk due to its bias towards the sample mean, or to the cluster means if it is a multimodal distribution. The set of subsamples $\mathbb{S}$ (Equation \ref{eq:set_subsample}) under-represents the outlier observations (over-represents the central observations), which biases the synthetic sample. This bias could be desirable because of the reason explained above. This section establishes a formal result showing that $\mathbb{S}$ under-represents the outlier observations, and hence disclosure risks driven by outliers are minimized.

\begin{definition}
  An outlier $o$ is defined as an observation that is \textbf{sufficiently} distant from the majority of other observations, such that 
  \begin{equation}\label{eq:outlier-equation}
    D_o > T,
  \end{equation}
  where $$D_i = \sum_{m=2}^{k} d(\bm {x}_i, \bm {x}_j)$$ and $T$ is a constant threshold defining outliers.  
\end{definition}

Note that the formally defining an outlier is not straightforward, and I presume that $T$ will be chosen such that the outliers will have a $D_o$ that is larger than the majority of other observations $D_i$, e.g., larger than 99\% of other observations.

\begin{theorem}\label{th:outlier}
Let $Z_i\in\{0,1,...,n\}$ be a random integer variable representing the frequency of observation $i$'s appearance in the set of subsamples \eqref{eq:set_subsample}. Assuming $E[Z_i|D_i]$ is a monotonic function of $D_i$, $$E[Z_i|D_i = D_o]<E[Z_i|D_i = D_c],$$where $o$ and $c$ denote outlier and central observations, respectively. 
\end{theorem}

\begin{proof}
  See Appendix \ref{sec:outlier-proof}.
\end{proof}

Theorem \ref{th:outlier} implies that the outlier observations are \textit{likely} to appear less frequently than other observations, specifically, compared to central observations that have lower $D_i$ values than outliers. Hence, we conclude that outliers are under-represented in the set of subsamples $\mathbb{S}$. 

\begin{remark}\label{rem:outlier-exclusion}
  While it is difficult to generalize the impact of $k$ on bias, LR with $k$NN can be further modified to increase the propensity to trim outliers. For example, the analyst can choose to exclude $i$ from the set of neighbors (Need to change the definition of Equation \eqref{eq:m-closest}). Consequently, outliers will not only appear less frequently in the set of subsamples $\mathbb{S}$ but could also be \textit{excluded} from the set of subsamples $\mathbb{S}$.
\end{remark}

\begin{remark}\label{rem:outlier-weighting}
  Another way to further underrepresent the outlier observations in the synthetic sample is estimating local distributions with distance-weighted observations. For each subsample, the distant observations will be penalized in the estimation of the distribution parameters. This approach will further bias the synthetic sample towards the sample mean or cluster means. 
\end{remark}

\begin{remark}\label{rem:outlier-sensitivity}
  Beyond the purpose of this article, Theorem \ref{th:outlier} suggests another practical use for LR with $k$NN. This method can serve as a sensitivity analysis tool, particularly in small samples where outliers can disproportionately impact statistical results. By running an analysis on both the original sample and a synthetic sample, analysts can compare results. Significant deviations in the outcomes could be driven by the sensitivity to outlier observations.
\end{remark}

\subsection{Discrete and Censored Values}

One of the challenges in estimating a multivariate distribution is often handling mixed-type variables, specifically continuous and discrete variables. This process can require complex and computationally heavy procedures, but it is feasible as demonstrated by previous research works such as \citep{onken2016mixed, ziegler2019latent, tran2019discrete, kamthe2021copula}.

Alternatively, one can synthesize discrete variables from continuous distributions and stochastically round the synthetic values as follows: 

\begin{equation}
  \tilde{ x}^{rounded} _{l,i}  = \begin{cases}
    \lfloor \tilde{ x} _{l,i} \rfloor & \text{with probability }  \tilde{ x} _{l,i}  - \lfloor \tilde{ x} _{l,i} \rfloor,\\ 
    \lceil \tilde{ x} _{l,i}\rceil & \text{with probability } \lceil \tilde{ x} _{l,i} \rceil - \tilde{ x} _{l,i}, \\ 
  \end{cases}
\end{equation}

where $\tilde{ x}_{l,i}$ is the $l^{th}$ covariate and $i^{th}$ synthetic observation. Implementing such a method may compromise the validity of assumption \ref{a:distribution}. Nevertheless, our simulation results indicate that the resulting bias is often minor and may be disregarded, especially when the analyst faces computational limitations.

\subsection{Missing Values}

The presence of missing values is a common issue in real-world datasets. The proposed method does not naturally handle missing values, and missing values should be imputed prior to the synthesis process. However, the analyst may opt to have missing values in the synthetic data as in the real data. In that case, missing values can be imputed, for example, with the MICE algorithm \citep{van1999flexible} while retaining the missingness indicators of each variable as a separate variable. The missingness indicators for each variable can be treated as an additional variable in the vector $\bm x _i$ and synthesized. Based on the synthesized missingness indicators (the missingness propensities), missing values can be generated stochastically.

\subsection{Identical Neighbors}\label{sec:identical}

The analyst can encounter cases where $k$ neighbors (or cluster $l$ with size $n_l$) are identical for certain discrete variables. Hence, estimating zero-variance distributions will be redundant. Nevertheless, this is not necessarily a concern if $k$ ($n_l$) is ``sufficiently large''. If there are identical neighbor variables, then the synthetic values will be created from a stratified subsample. Given that $k$ ($n_l$) is ``sufficiently large'', there will not be any disclosure risk because there are many entities in the population with similar characteristics.

\section{Simulations}\label{sec:simulations}

This section evaluates the performance of the Local Resampler (LR) algorithm in replicating complex data distributions. We conduct two simulation experiments, each designed to test LR's ability to capture challenging distributional features such as multimodality, skewness, and non-convex support.

\subsection{Two-Rings Simulation}

In the first simulation, I generate data forming two concentric rings, with one ring surrounding the other. Replicating such a distribution is challenging due to the non-convex support (the empty space between the rings), which results in a multimodal distribution.\footnote{The two-ring concept is adapted from \citet{kamthe2021copula}.}

The data generating process is as follows:

\begin{align*}
    u_{j,i} &\sim \text{Normal}(0, 1), \\
    x_i &\sim r_i \times \cos{a_i} + u_{1,i}, \\
    y_i &\sim 0.5 x_i - 0.05 x_i^2 + r_i \times \sin{a_i} + u_{2,i},
\end{align*}

where $r_i$ takes values 8 or 20 with equal probability, and $a_i$ is drawn from a uniform distribution between 0 and $2\pi$.

\begin{figure}[!ht]
  \includegraphics[width=\textwidth]{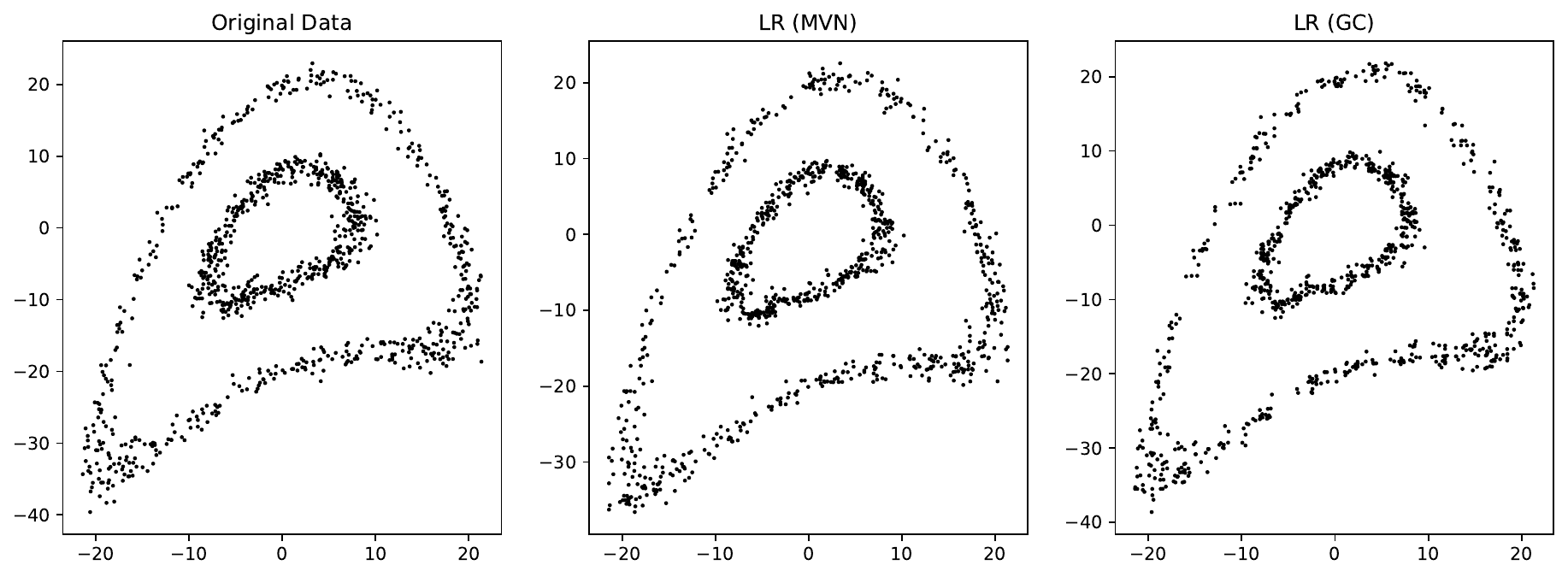}
  \caption{Scatter plots of the original (left) and synthesized samples.}\label{f:bivariate}
  \raggedright
  \scriptsize \textbf{Note:} The scatter plot on the left shows the original sample. The second scatter plot shows the synthetic sample created using $k$NN with a multivariate normal distribution (MVN). The third scatter plot shows the synthetic sample generated using $k$NN with Gaussian Copula (GC).
\end{figure}

The sample size is 1,000, and the number of neighbors ($k$) for the $k$NN algorithm is set to 15. I estimate local distributions using both the multivariate normal distribution (MVN) and the Gaussian Copula (GC). Figure \ref{f:bivariate} presents scatter plots comparing the original sample with synthetic samples generated using LR with $k$NN-MVN and $k$NN-GC. Both methods replicate the original sample accurately, effectively capturing the non-convex support and multimodality. Notably, no significant differences are observed between using MVN or GC in terms of replicating the sample structure.

\subsection{Clustered Distribution Simulation}

The second simulation involves generating a distribution characterized by clustered observations located at different positions within the sample space:

\begin{align*}
x_i &\sim \text{Beta}(0.1, 0.1),\\
y_i &\sim \text{Beta}(0.1, 0.5),\\
z_i &\sim 10 y_i + u_i,
\end{align*}

where $u_i$ is drawn from a standard normal distribution. The resulting distribution is highly skewed, with distinct clusters primarily located at the boundaries of the sample space. More importantly, this data generating process produces outlier observations that are not detectable in the marginal distributions of the $x$, $y$ and $z$ variables. Hence, this data generating process aims to illustrate the implications of Theorem \ref{th:outlier}.

\begin{figure}[!ht]
  \centering
  \includegraphics[width=\textwidth]{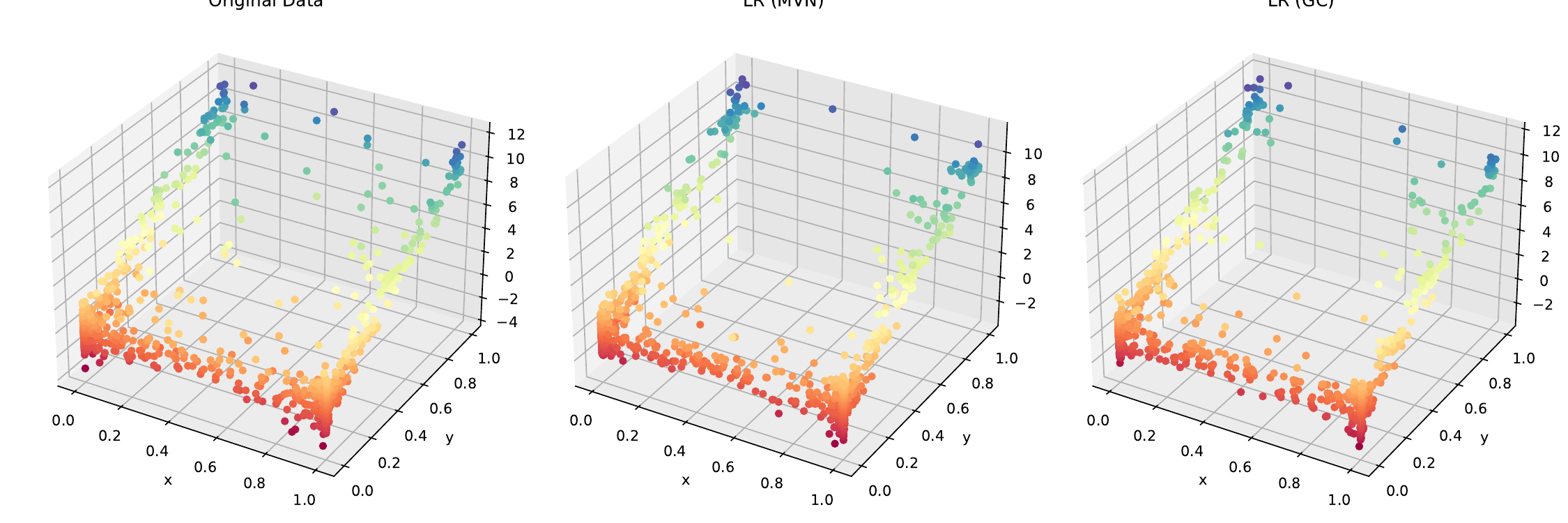}
  \caption{3D scatter plots of the original and synthetic samples.}\label{f:trivariate}
  \raggedright
  \scriptsize \textbf{Note:} The scatter plot on the top left shows the original sample. The middle scatter plot shows the synthetic sample created using $k$NN with a multivariate normal distribution (MVN). The third scatter plot shows the synthetic sample generated using $k$NN with Gaussian Copula (GC).
\end{figure}

Figure \ref{f:trivariate} (left) displays the original data distribution, and the middle and right panels illustrate synthetic data generated with LR using $k$NN-MVN and $k$NN-GC, respectively.

The results demonstrate that LR with $k$NN-MVN and $k$NN-GC accurately replicates the original sample, effectively capturing the multimodal and skewed distribution while maintaining the convex support. Notably, the central observations in the original dataset represent outliers that cannot be easily identified by the analyst through marginal distributions alone, as their low probability densities become evident only in the joint distribution. The LR method inherently reduces the presence of such outliers by underrepresenting them during the resampling process. As a result, the synthetic datasets generated by LR are less likely to contain these outlier observations, which effectively mitigates disclosure risks, as formally established by Theorem \ref{th:outlier}.

\subsection{Comparison with Existing Methods}

\begin{figure}[!ht]
  \centering
  \includegraphics[width=0.8\textwidth]{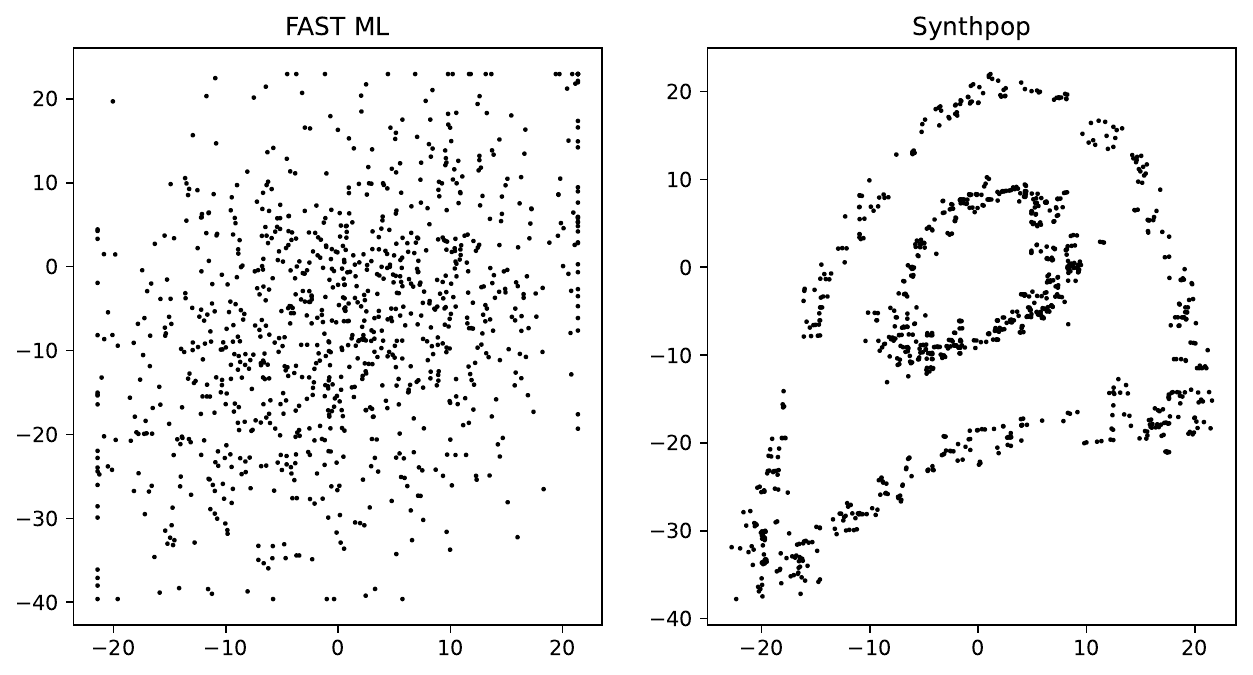}
  \caption{Scatter plots of the first simulation with FAST ML and synthpop.}
  \label{f:synthpop_circular}
  \includegraphics[width=0.8\textwidth]{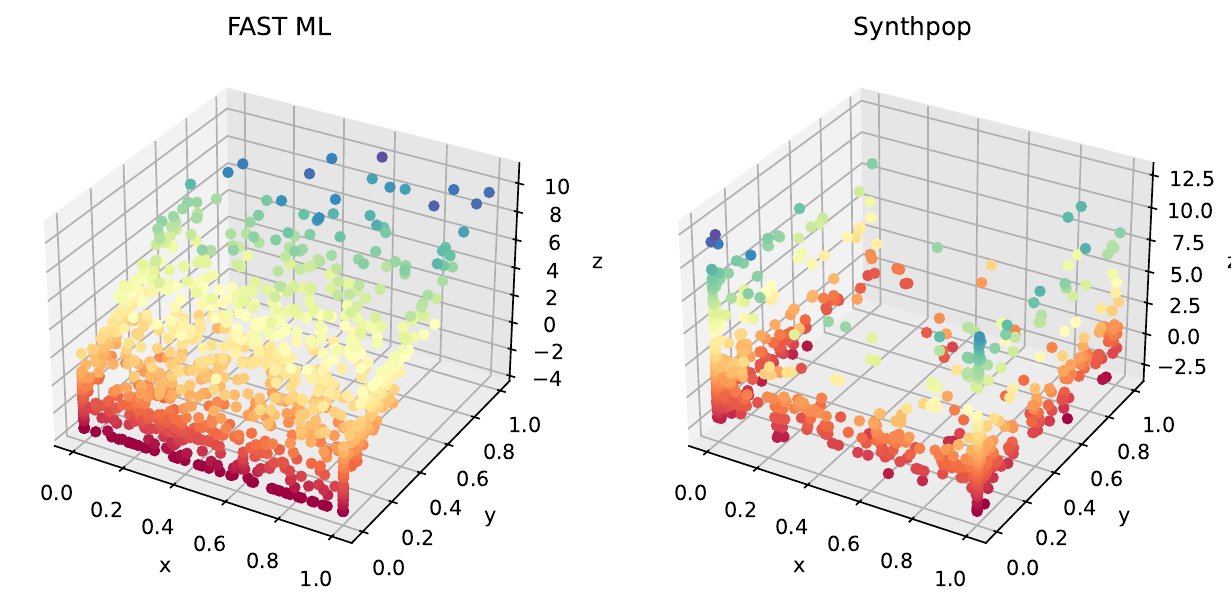}
  \caption{3D scatter plots of the second simulation with FAST ML and synthpop.}
  \label{f:synthpop_trivariate}
\end{figure}

To further assess the performance of LR, we compared it with the \textbf{FAST ML} and \textbf{synthpop} approaches using the same artificially generated datasets. Figures \ref{f:synthpop_circular} and \ref{f:synthpop_trivariate} present scatter plots of synthetic samples generated by these methods.

\textbf{synthpop} performed relatively well in replicating the circular distributions but struggled to replicate the clustered, trivariate distribution accurately. \textbf{FAST ML}, on the other hand, failed to effectively replicate both distributions, particularly in capturing the complex relationships between variables.

The results highlight that LR outperforms both \textbf{FAST ML} and \textbf{synthpop}, particularly when dealing with challenging artificial datasets characterized by multimodality, skewness, and non-convex support. The simulations confirm that LR is highly effective in replicating complex distributions, accurately preserving key structural features.

It is important to note that the data generating processes used in these simulations were intentionally designed to be (1) difficult to replicate, incorporating features and (2) to contain outlier observations that are not detectable in the marginal distributions to show the implications of Theorem \ref{th:outlier}. In contrast, real datasets are often less challenging, particularly if they are not high-dimensional. Appendix \ref{sec:realdata} demonstrates the application of LR on real datasets and provides a comparison of multivariate regression analysis results using LR, \textbf{FAST ML}, and \textbf{synthpop}. While LR shows slightly better performance overall, \textbf{synthpop} achieves comparable results to LR.

\section{Conclusion} \label{sec:conclusion}

This paper addresses a critical challenge in the responsible use of sensitive datasets by using synthetic data: the disclosure risks posed by outliers. LR offers a compelling solution. LR's ability to accurately replicate complex distributions while underrepresenting outliers advances statistical disclosure control. This inherent bias towards centrality automates the typically manual and limited process of outlier handling.

The policy implications of LR are significant. By providing a reliable, computationally efficient, and statistically robust method for generating synthetic data, LR can facilitate greater access to sensitive datasets for policy-relevant research. This access can inform evidence-based policymaking, particularly in data-driven fields like economics and healthcare. LR's adaptability to different subsampling methods (e.g., clustering), as shown in Appendix \ref{app:clustering}, allows for tailoring the privacy-utility trade-off to specific policy contexts and research objectives. Moreover, as shown by Remarks \ref{rem:outlier-exclusion} and \ref{rem:outlier-weighting}, LR can be further customized to enhance outlier handling.

While LR offers clear advantages, further research could explore its integration with other disclosure control techniques to enhance its adaptability across diverse datasets and policy domains. Additionally, investigating LR's performance under different sample sizes and distributional assumptions would further strengthen its practical applications. In conclusion, LR holds the potential to promote research using sensitive data. This work contributes to the development of robust data governance practices and policies, promoting ethical and impactful research. 

\section*{Acknowledgements}

I am deeply grateful to my PhD supervisors, Prof. Alicia N. Rambaldi and Dr. Christiern Rose, for their valuable support. I also greatly benefited from the guidance of Jonas Fooken, Yuanyuan Gu, and Metin Uyanik in preparing this work for submission.

\subsection*{Declaration of generative AI and AI-assisted technologies in the writing process.}

During the preparation of this work the author used Google Gemini and OpenAI GPT4 in order to enhance readibility of the article. After using these tools, the author reviewed and edited the content as needed and takes full responsibility for the content of the publication.

\bibliographystyle{apacite}
\bibliography{refbib}

\newpage

\appendix

\appendix 

\renewcommand{\thefigure}{A.\arabic{figure}}
\setcounter{figure}{0}

\renewcommand{\thetable}{A.\arabic{table}}
\setcounter{table}{0}

\renewcommand{\theequation}{A.\arabic{equation}}
\setcounter{equation}{0}

\section{The Trend of using ``Administrative Data'' for Research}\label{app:academic_publications}

\begin{figure}[!ht]
  \centering
  \includegraphics[width=0.8\textwidth]{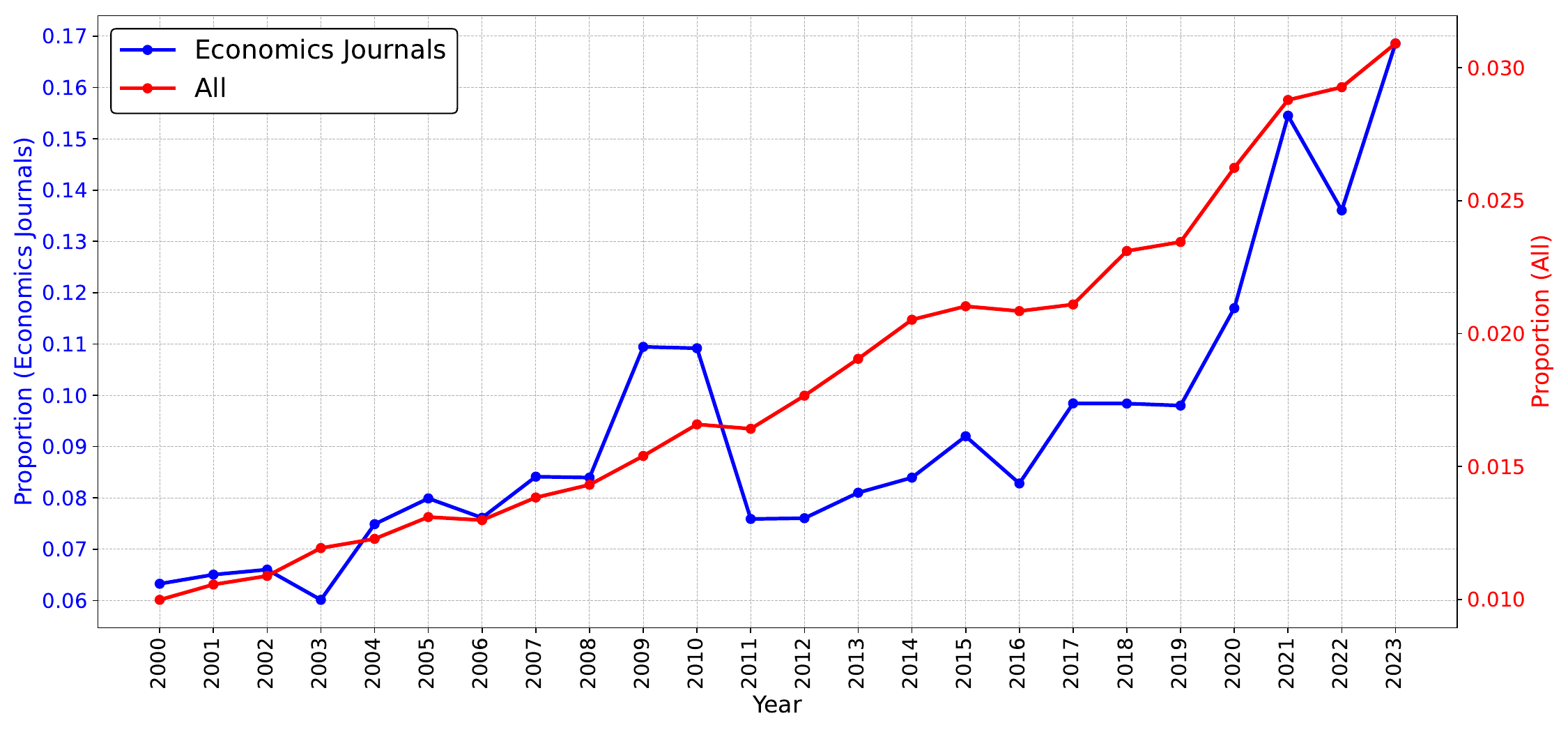}
  \caption{The Trend of using ``Administrative Data'' for Research}
  \label{fig:academic_publications}
\end{figure}

Figure \ref{fig:academic_publications} displays the proportion of search outcomes containing the term ``administrative data'' relative to the total number of publications for each corresponding year. Searches specific to Economics journals (blue curve) employed the ``Subject areas'' filter, selecting ``Economics, Econometrics, and Finance''. For a broader overview encompassing all journals (red curve), a more general search was carried out. To gauge total outcomes in Economics journals, a space `` '' served as the search parameter, while the letter ``a'' represented all journals. The y-axis on the left illustrates proportions for Economics journals, whereas the y-axis on the right is for all journals. Data for 2023 was sourced as of 22 October, and thus might not cover the entire year. 

\section{Discussions with Formal Assumptions}\label{sec:formal-assumptions}

I discuss the formal assumptions in the appendix as LR is a inherently biased approach. The nature of this bias is explained by Theorem \ref{th:outlier}, which is crucial for disclosure control. This section outlines the formal assumptions needed to understand the LR approach and examines how the distance-based subsampling influences the estimation process.

\begin{assumption}
  Observations in subsample $S_i$ are drawn from an unknown distribution $F_i(\bm{\alpha }_i)$ with parameters $\bm{\alpha }_i$. $\hat F_i(\hat{\bm{\alpha }_i})$ is an unbiased estimator of $F_i(\bm{\alpha }_i)$ for all $i\in S'$ where $\hat{\bm{\alpha }_i}$ is estimated with subsample $S_i$. \label{a:distribution}
\end{assumption}

Under the assumption \ref{a:distribution}, the target inference from the synthetic sample will be an unbiased estimator of the target inference from the original sample. Assumption \ref{a:distribution} is, in fact, a set of $n$ nested distributional assumptions and is likely to fail, at least, for some subsamples in $\mathbb{S}$. However, it is possible to test the validity of the distributional assumption for each subsample $S_j\in\mathbb{S}$ or use nonparametric distributions, if not unfeasible due to high dimensionality. As simulations in this paper shows, Assumption \ref{a:distribution} is less restrictive than assuming that the entire sample is drawn from a fixed parametric or nonparametric distribution. 

\begin{figure}[!ht]
  \centering
  \includegraphics[width=0.75\textwidth]{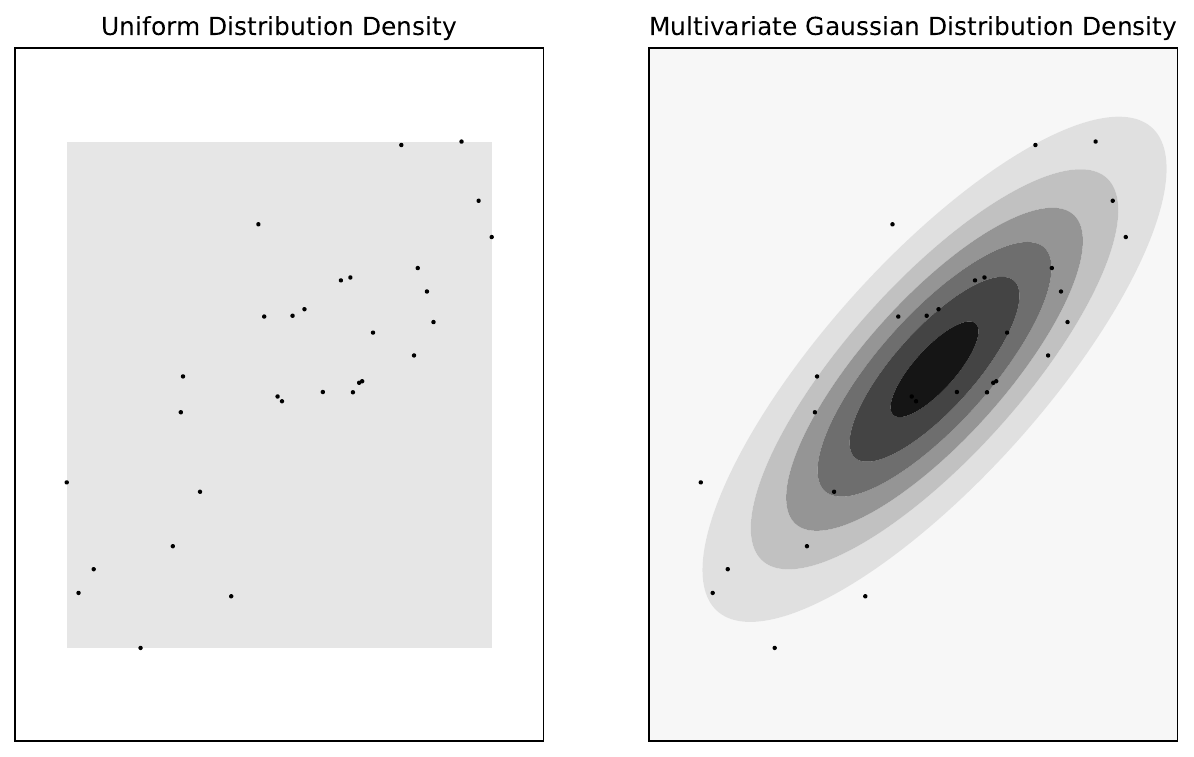}
  \caption{Estimated Probability Density Functions of Uniform (left) and Multivariate Normal (right) Distributions for a Small Sample}\label{f:uniform_vs_normal}
  \raggedright
  \scriptsize{\textbf{Note:} The darker shades indicate higher probability densities.}
\end{figure}

The choice of a distribution model can influence the quality of synthetic samples. Figure \ref{f:uniform_vs_normal} demonstrates a simple case where the probability density functions estimated using both uniform (as histograms do) and multivariate normal (MVN) distributions for a small sample (e.g., a subsample). The correlation visible in the figure might be overlooked if one were to estimate using a uniform distribution. However, as subsamples are defined based on their proximity to each other, the complexity of the sample distribution tends to diminish. Consequently, using a parametric distribution, such as MVN or Gaussian Copula, is often sufficient with LR algorithm.

$k$NN is generally assumed to have constant $k$ for the asymptotic consistency as a predictor. The convergence rate of the estimator depends on $p$, which is known to suffer from the ``curse of dimensionality''. However, the proposed methodology uses $k$NN only to define subsamples and $k$NN is not required to make asymptotically consistent predictions. Therefore, it is assumed that $k$ increases at the same rate as the sample size.

\begin{assumption}
  $k/n\to a$ as $n\to \infty$ where $a \in (0,1)$. \label{a:k_rate}
\end{assumption}

Assumptions \ref{a:distribution} and \ref{a:k_rate} imply that the estimated distribution parameters should converge to the true distributions fast. Under some parametric assumptions (e.g., MVN), the estimated distributions could converge at rate $O_p(n^{-1/2})$. 

As $k$ increases, the bias of the estimated distribution will increase and Assumption \ref{a:distribution} will be more restrictive. Consider the following scenarios:

\begin{description}
  \item[$k=1$:] (Although we assumed $k>1$ earlier) It implies that estimated distributions will have zero variance. Therefore, the original sample will be copied and resampled with replacement. In other words, it becomes a bootstrapping sample with replacement.  
  \item[$k=2$:] Values will be drawn between/around two points. For example, if a uniform distribution is estimated we will draw samples from a line (if it is a sample with one variable), a rectangle (2 variables), a cube (3 variables), or a tesseract (4 variables) with equal probability. 
  \item[$k=n$:] The entire sample will be used to estimate a distribution. This is the standard practice in the literature, where the entire distribution is fitted to a specific distribution or model.
\end{description}

It is not trivial to make generalizations about the effect of $k$ on the synthetic sample. Given that this paper employs the multivariate normal distribution and Gaussian copulas, it is recommended to maintain $k>10$. Nevertheless, the choice will depend on the sample size and the dimensionality.

\section{Local Resampler Algorithm with Clustering}\label{app:clustering}

An alternative to $k$NN is using clustering algorithms to define subsamples. Then, distributions are estimated for each cluster and synthetic values are created from the estimated distributions. The main challenges are choosing the number of clusters and using clustering constraints. For this purpose, we will employ a constrained K-Means algorithm proposed by \citet{bradley2000constrained}.

Let $C_l$ be cluster $l\in\{1,2,...,L \}$ with size $n_l$ defined by the K-Means algorithm. The estimated distribution for cluster $C_l$ is denoted by $\hat F_{C_l}$. Let $\mathbb{C}$ be the set of the estimated distributions of clusters, 

\begin{equation}
  \mathbb{C} = \left \{ \hat F_{C_1}, \hat F_{C_2}, ..., \hat F_{C_{L}}    \right \}.\label{eq:set_of_dist2}
\end{equation}

The analyst can now create synthetic values from the estimated distributions, with size $n'$ by resampling $\mathbb{C}$ with replacement. Each $\hat F_{C_l}$ will be drawn with a probability proportional to the size of cluster $l$. Let $\mathbb{C'}$ be the resampled set of the estimated distributions with size $n'$,

\begin{equation}
  \mathbb{C'} = \left \{ \hat F_{C_{(1)}}, \hat F_{C_{(2)}}, ..., \hat F_{C_{(n')}}    \right \}.\label{eq:resampled_set_of_dist}
\end{equation}

The analyst may simply include $\hat F_{C_l}$ in $\mathbb{C'}$, $n_l^{th}$ times instead of using the probability proportional to the size of cluster $l$. The synthetic values will be drawn from $\mathbb{C'}$,

\begin{equation}
  \left \{ \tilde{\bm x} _i\right \}_{i=1} ^{n'} = \left \{ \tilde{\bm x _{1}}, \tilde{\bm x _{2}}, ..., \tilde{\bm x _{n'}}\right \} \sim \mathbb{C'} .\label{eq:draw_Sample_cluster}
\end{equation}

\begin{algorithm}[!ht]
  \KwData{$\bm x$, of dimension $n\times p$ where $n$ is the sample size and $p$ is the number of variables.\tcp*{$\bm x_i$ is the $i^{th}$ observation from data}}
  \KwIn{$k$: the number of neighbors; $\theta$: parameters for K-Means algorithm; $n'$: the required synthetic sample size.}
  \KwOut{$\tilde{\bm x}\leftarrow n'\times p$\tcp*{A synthetic dataset}}
  Assign each $i\in \{1,2,...,n\}$ to $L$ clusters with the constrained K-Means algorithm with parameters $\theta$\;
  $C_l\leftarrow $ Cluster $l\in\{1,2,...,L \}$ with size $n_l$\;
  \tcc{Estimating distributions for each cluster}
  \For{$l\in \{1,2,...,L\}$}{
    $\hat F_{C_l} \leftarrow $ Estimated distribution from cluster $C_l$ \;
  }
  \tcc{Draw synthetic values from the estimated distributions with required size $n'$}
  \For{$i\in \{1,2,...,n'\}$}{
    $\bm {F}_i\leftarrow $ Randomly draw an estimated distribution from $\{\hat F_{C_1}, \hat F_{C_2},..., \hat F_{C_L}\}$ with respective probabilities $\{ \frac{n_1}{n}, \frac{n_2}{n},..., \frac{n_L}{n}\}$\;
    $\tilde{\bm x _{i}}\leftarrow $ Draw a synthetic value from $\bm{F}_i$ \;
  }
  \KwRet{$\tilde{\bm x} \leftarrow n'\times p$}
  \caption{Creating Synthetic Data with Clustering Algorithm}
  \label{alg:cluster}
\end{algorithm}

Algorithm \ref{alg:cluster} summarizes the steps of the proposed methodology in this section. Henceforth, this algorithm shall be referred to as the cluster local resampler (C-LR).

\subsection{Assumptions for Clustering Algorithm}

Assumptions for C-LR can be defined similarly to those for $k$NN-LR with slight modifications. 

\begin{assumptionp}{\ref*{a:distribution}$'$}
  Observations in cluster $C_l$ are drawn from an unknown distribution $F_l(\bm{\alpha }_l)$ with parameters $\bm{\alpha }_l$. $\hat F_l(\hat{\bm{\alpha }_l})$ is an unbiased estimator of $F_l(\bm{\alpha }_l)$ for all $l\in \{1,2,...,L\}$ where $\hat{\bm{\alpha }_l}$ is estimated with cluster sample $C_l$. \label{a:nl_distribution} 
\end{assumptionp}

\begin{assumptionp}{\ref*{a:k_rate}$'$}
  $n_l/ n \to C$ for all $l \in \{1,2, ...,L\}$as $n\to \infty$ where $C \in (0,1)$.\label{a:nl_rate} 
\end{assumptionp}

The asymptotic consequences of assumptions \ref{a:nl_distribution} and \ref{a:nl_rate} resemble those of assumptions \ref{a:distribution} and \ref{a:k_rate}. However, $k$NN-LR differs from C-LR in an important detail, which is discussed in Section \ref{sec:trimming}.

\subsection{Simulation Results with the Clustering Algorithm}

\begin{figure}[!ht]
  \centering
  \includegraphics[width=0.8\textwidth]{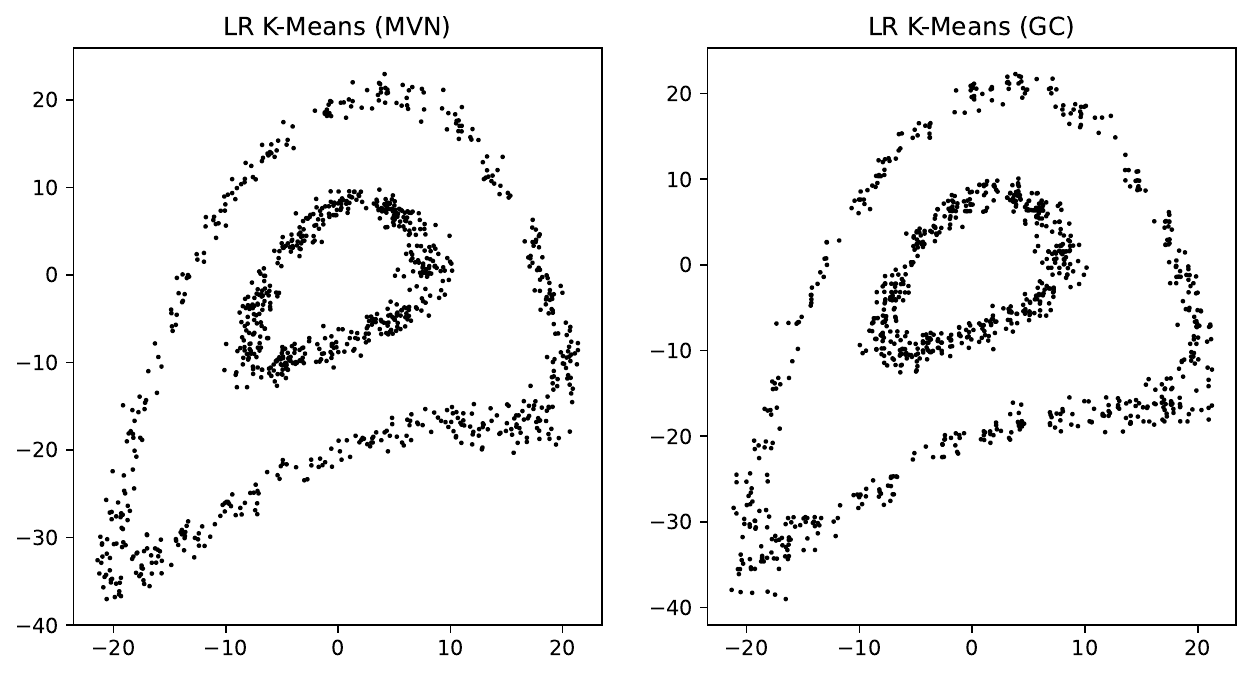}
  \caption{Scatter plots of the original (left) and synthesized samples.}\label{f:bivariate_kmeans}
  \raggedright
  \scriptsize
  \textbf{Notes:} This application replicates the results in Figure \ref{f:bivariate} using the clustering algorithm.
\end{figure}

\begin{figure}[!ht]
  \centering
  \includegraphics[width=0.8\textwidth]{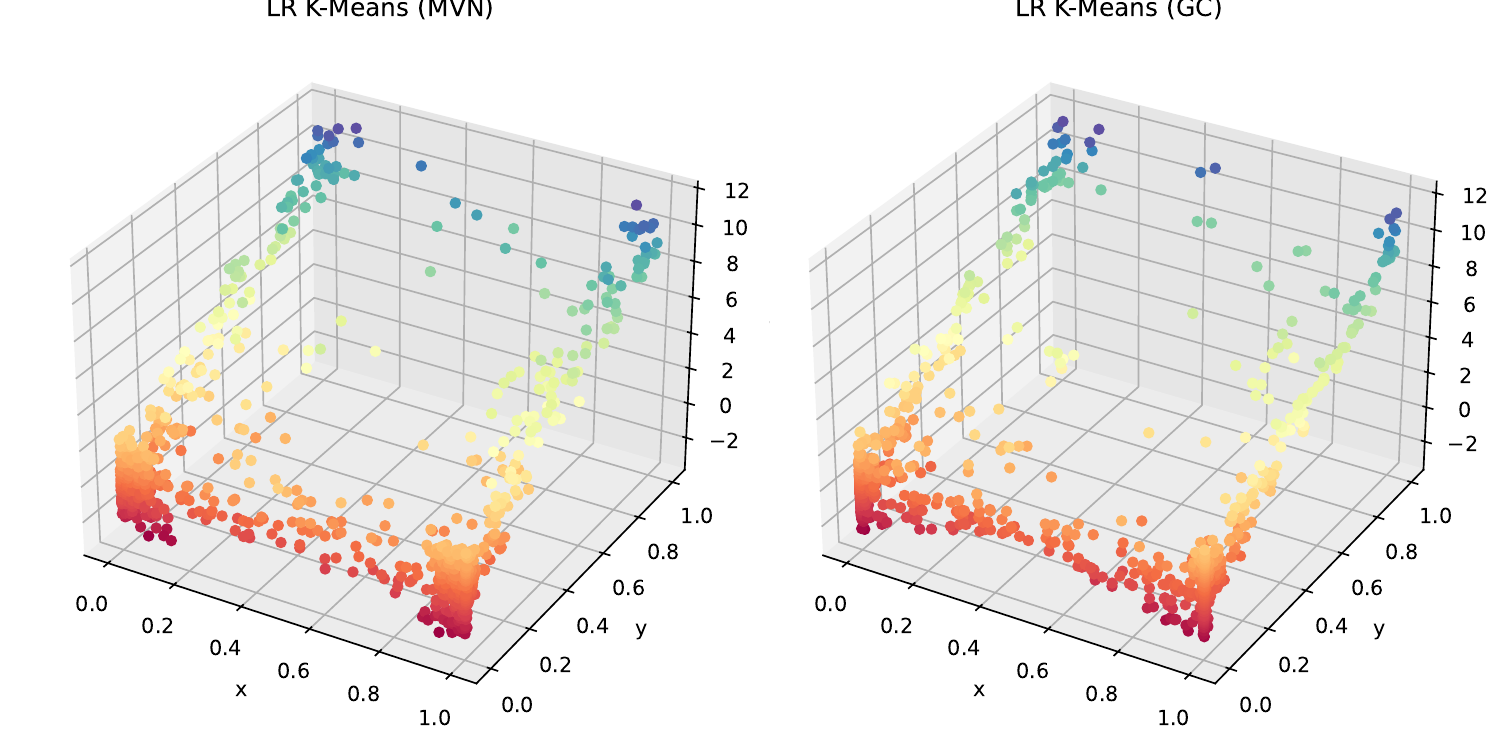}
  \caption{Scatter plots of synthetic samples created with K-Means algorithm using multivariate normal distribution (left) and Gaussian Copula (right).}\label{f:trivariate_kmeans}
  \label{f:kmeans-trivariate}
  \raggedright
  \scriptsize
  \textbf{Notes:} This application replicates the results in Figure \ref{f:trivariate} using K-Means algorithm. 
\end{figure}

I use a constrained K-means algorithm \citep{bradley2000constrained} with a cluster size of 30 and with minimum cluster size restricted to 10. Figures \ref{f:bivariate_kmeans} and \ref{f:trivariate_kmeans} demonstrates the synthetic samples created with the clustering algorithm. The results are similar to those obtained with $k$NN-LR, respectively, in Figures \ref{f:bivariate} and \ref{f:trivariate}. Results show no visible difference between $k$NN and clustering algorithms, except in Figure \ref{f:trivariate_kmeans} outlier observations appear as frequently as in the original sample unlike as in Figure \ref{f:trivariate}.

\section{Proof of Theorem \ref{th:outlier}}\label{sec:outlier-proof}

The theorem makes the assumption that $E[Z_i|D_i]$ is a monotonic function of $D_i$, which is not necessarily correct, especially with the complex distributions. However, this assumption is not concerning because we assume that there is sufficient disparity between outliers and central observations in terms of $D_i$. Given that the theorem's main purpose is to compare an outlier $(o)$ to central observations $(c)$, where the disparity between $D_o$ and $D_c$ is expected to be large, the monotonicity is not a strong assumption. For example, $D_o$ could be the largest $o\in \mathbb{S}$ and $D_c$ could be an observation within the smallest $25^{th}$ percentile of $\{D_i \}_{i=1}^n$. 

Now, we use the following lemma to prove Theorem \ref{th:outlier}.

\begin{lemma}\label{lem:negative-correlation}
  $E[Z_i|D_i]$ is decreasing in $D_i$ \eqref{eq:outlier-equation}.
\end{lemma}

\begin{proof}  
  \textbf{Relationship between distance and subsample appearance:} Consider two observations, $i$ and $j$, and their respective distances to all other observations. Given the definition of a subsample set $S_j$, observation $i$ will appear in $S_j$ if $i$ is among the $k$ nearest neighbors to $j$. The probability $\mathbb{P}(i \in S_j)$ is therefore inversely related to the rank of $i$ when all observations are sorted according to their distance from $j$. 

  For simplicity, let's denote $r(i, j)$ as the rank of observation $i$ when observations are sorted by increasing distance from $j$. We can then infer: $$I\left ( i \in S_j \right ) = I\left ( r(i, j)\leq K \right  ),$$ where $I(.)$ is an indicator function that takes the value 1 if the statement is true and 0 otherwise. 

  \textbf{Aggregated effect over all subsamples:} Given the definition of $D_i$, which measures the cumulative distance of $i$ to its $k$ nearest neighbors, we define an average rank of $i$ across all observations, $$R_i = \frac{1}{n} \sum_{j=2}^k r(i, i(j)).$$Given this relationship, as $D_i$ increases (i.e., an observation is on average farther from other observations), the average rank $R_i$ should also increase. This establishes:$$R_i \propto D_i.$$Consequently, the expectation of $Z_i$ (the frequency of $i$ in all subsamples) is inversely related to $R_i$ and, thus, inversely related to $D_i$. Formally:$$E[Z_i|D_i] \propto \frac{1}{R_i} \propto \frac{1}{D_i},$$ hence, $E[Z_i|D_i]$ is decreasing in $D_i$.
\end{proof}

Having Lemma \ref{lem:negative-correlation} established, the proof of Theorem \ref{th:outlier} is trivial. Outliers, by definition, have greater distances from the majority of observations, $$D_o > D_c.$$ Using the established relationship from Lemma \ref{lem:negative-correlation}: $$E[Z_i|D_i=D_o] \propto \frac{1}{D_o} < \frac{1}{D_c} \propto E[Z_i|D_i=D_c].$$
Thus, $$E[Z_i|D_i=D_o] < E[Z_i|D_i=D_c].$$

\section{Application with a Real World Dataset} \label{sec:realdata}

This section evaluates the performance of LR algorithms in synthesizing tabular data, using a comparison with established methodologies on a real-world dataset. The analysis involves comparing the marginal distribution of each variable in the actual dataset and the estimated coefficients from a multivariate linear regression, applied to both the original and the synthesized datasets. The synthesized dataset comprises 20,640 observations from housing data in California districts, sourced from the 1990 US Census. It includes variables such as median house prices, median income, average age of the house, average number of bedrooms, population, and average household size in the Census districts \citep{pace1997sparse}.

I set $k=15$ for the $k$NN algorithm and established a minimum cluster size constraint of 50 for the K-means algorithm. Table \ref{t:descriptives} displays the descriptive statistics of the original and the synthetic samples created with LRs.\footnote{I excluded $k$NN-LR with the Gaussian Copula due to the ease of visualization and the fact that its results were almost identical to those obtained with MVN.} The synthesized dataset's average and median values are very close to the original sample, but the standard errors are smaller than those of the original sample for all variables, due to the result in Theorem \ref{th:outlier}. 

\begin{table}[!hb]
  \centering
\caption{Descriptive statistics of the original and synthetic samples.}
\label{t:descriptives}
  \resizebox{\textwidth}{!}{%
\begin{tabular}{lrrrrrrrrr}
    \toprule
    & \multicolumn{3}{c}{\textbf{Original}} & \multicolumn{3}{c}{\textbf{Synthetic: $k$NN-LR (MVN)}} & \multicolumn{3}{c}{\textbf{Synthetic: $k$NN-LR (GC)}} \\
     & \textit{Average} & Std.Dev & Median & \textit{Average} & Std.Dev & Median & \textit{Average} & Std.Dev & Median \\
    \midrule
    \textit{MedInc} & 3.87 & 1.90 & 3.54 & 3.87 & 1.86 & 3.53 & 3.87 & 1.86 & 3.53 \\
    \textit{HouseAge} & 28.64 & 12.59 & 29.00 & 28.67 & 12.37 & 28.81 & 28.69 & 12.41 & 29.00 \\
    \textit{AveRooms} & 5.43 & 2.47 & 5.23 & 5.42 & 2.22 & 5.23 & 5.42 & 2.15 & 5.22 \\
    \textit{AveBedrms} & 1.10 & 0.47 & 1.05 & 1.09 & 0.40 & 1.05 & 1.09 & 0.42 & 1.05 \\
    \textit{Population} & 1425.48 & 1132.46 & 1166.00 & 1426.18 & 1068.25 & 1159.29 & 1422.68 & 1046.09 & 1156.03 \\
    \textit{AveOccup} & 3.07 & 10.39 & 2.82 & 2.94 & 4.50 & 2.85 & 2.89 & 0.86 & 2.81 \\
    \textit{MedHouseVal} & 2.07 & 1.15 & 1.80 & 2.06 & 1.14 & 1.79 & 2.06 & 1.14 & 1.78 \\
    \bottomrule
\end{tabular}%
}

  \scriptsize
  \begin{tablenotes}
      \item \textbf{Notes:} The descriptive statistics at the top-right table is created with $k$NN algorithm using multivariate normal distribution (MVN). The synthetic samples at the bottom two tables were created with the clustering algorithm using multivariate normal distribution (MVN) and Gaussian Copula (GC).
  \end{tablenotes}
\end{table}

\begin{figure}[!ht]
  \centering
    \includegraphics[width=0.7\textwidth]{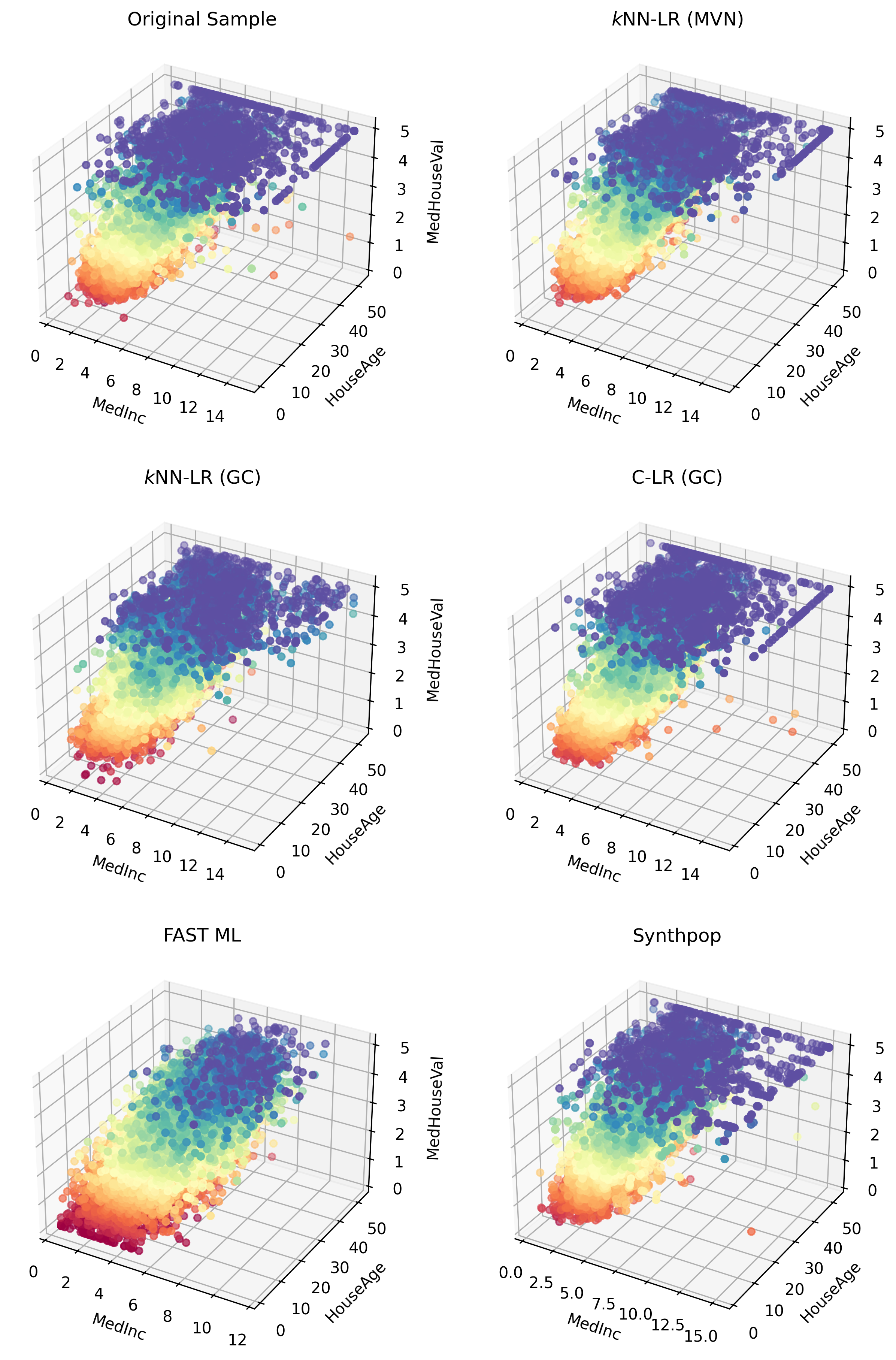}
    \caption{Scatter plots of original (upper left), local covariance (upper right), \textbf{FAST ML} (bottom left) and synthpop (bottom right) samples. The scatter plots show median income (x-axis), average house (y-axis) age, and median house prices (z-axis).}
    \label{f:compare_methods}
  \end{figure} 

\begin{figure}[!ht]
  \centering
  \includegraphics[height=0.82\textheight, width=0.9\textwidth]{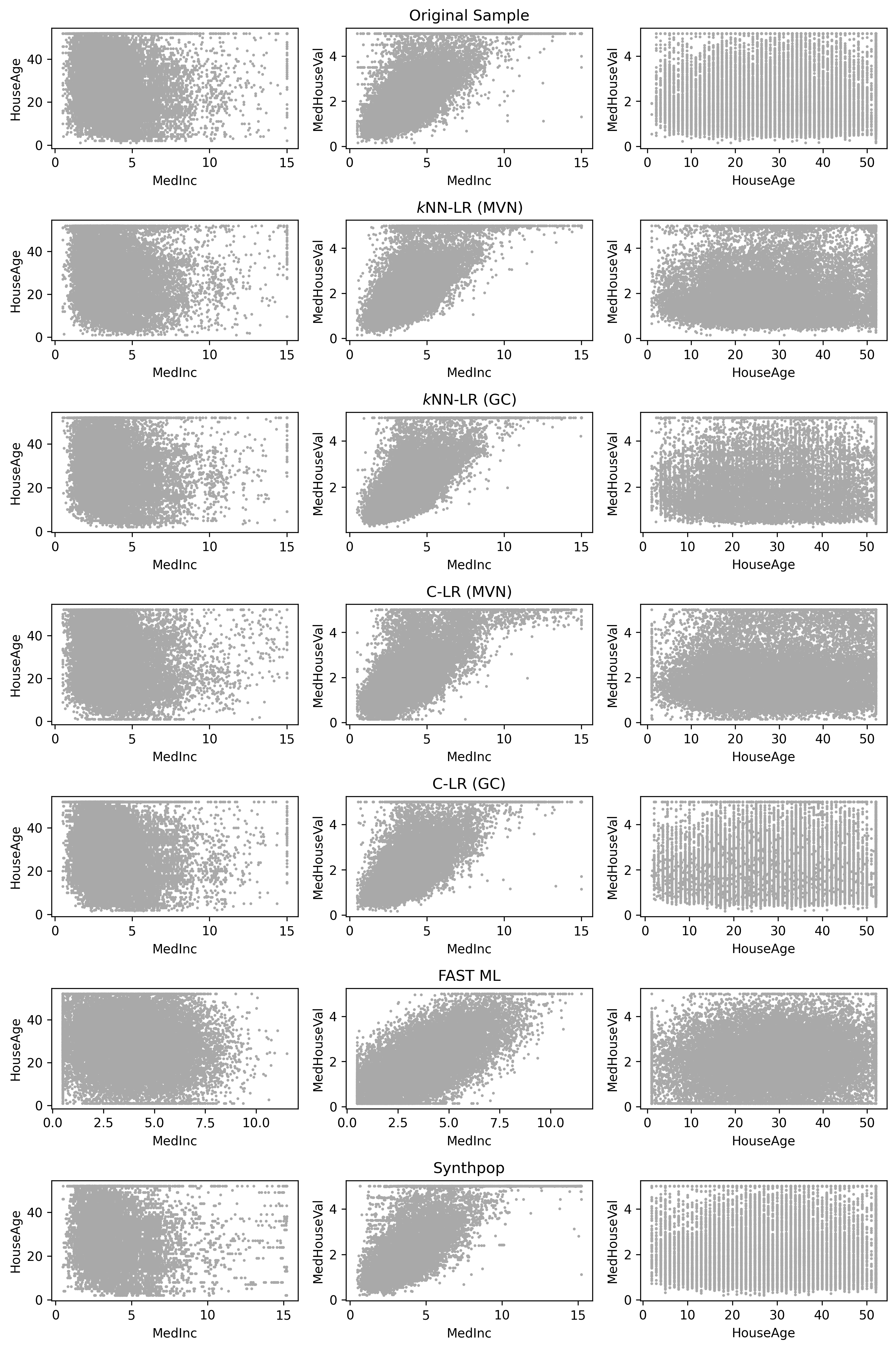}
  \caption{2D Scatter plot of Figure \ref{f:compare_methods}.} \label{f:compare_methods_2d}
  \raggedright
  \scriptsize
  \textbf{Notes:} Row 1: Original sample, Row 2: LR with $k$NN-MVN, Row 3: LR with $k$NN with Gaussian Copula, Row 4: LR with Clustering-MVN, Row 5: LR with Clustering-GC, Row 6: \textbf{FAST ML}, Row 7: \textbf{synthpop}. The discreteness of some variables in the original samples was only accounted by synthpop and LR with Gaussian Copula distributions. 
\end{figure}

The performance of LRs is now evaluated in comparison to other commonly utilized methods, specifically using the \textbf{SDV} software \citep{SDVpackage} and \textbf{synthpop} \citep{nowok2016synthpop}. Both software packages are popular choices. I selected a preset model named \textbf{FAST ML} in \textbf{SDV} and used the Classification and Regression Trees (CART) model with \textbf{synthpop}.\footnote{CART (Classification and Regression Trees) is widely employed both for imputing missing values (e.g., \citet{stekhoven2012missforest}) and generating synthetic data (e.g., \citet{drechsler2011empirical}) due to its robustness and flexibility. The minimum number of observations in any terminal node was set to 5 to prevent CART from overfitting. Additionally, the variables were smoothed with spline functions. This particular configuration was adopted following \citet{nowok2016synthpop}.} Figure \ref{f:compare_methods} presents scatter plots of three variables from the dataset. The synthetic samples produced by LRs and \textbf{synthpop} are visually more accurate compared to those produced by \textbf{FAST ML}. However, visual inspection is challenging with a 3D scatter plot. The distribution of each pair can be more effectively inspected in Figure \ref{f:compare_methods_2d}.

To test the accuracy of replicating marginal distributions, the Kolmogorov-Smirnov Distance statistics for each variable were calculated \citep{massey1951kolmogorov}. This metric ranges from 0 to 1, with smaller distances (disparity) indicating greater similarity in the empirical cumulative distribution functions. The average distance metrics for all variables were approximately 2\% for $k$NN-LR with MVN and GC, 16\% for the \textbf{FAST ML} method, and 1\% for the \textbf{synthpop} method. These findings align with the visual inspections from Figures \ref{f:compare_methods} and \ref{f:compare_methods_2d}. 

I now proceed to estimate multivariate regression models to compare the estimated coefficients when using the original and synthetic samples. Specifically, I estimate a simple hedonic regression for the median house value in respective census blocks. Table \ref{t:reg_table} shows the estimated coefficients of the following equation,\footnote{The regressors are median income, house age, average number of rooms per household, average number of bedrooms per household, population, and average number of household members, respectively.}

\begin{align}
  MedHouseVal_i =& \beta_0 + \beta_1 MedInc_i + \beta_2 HouseAge_i + \beta_3 AveRooms_i \notag \\
  &+ \beta_4 AveBedrms_i + \beta_5 Population_i + \beta_6 AveOccup_i + \epsilon_i. \label{e:reg}
\end{align}

\begin{table}[!hb]
  \caption{Regression coefficients estimated with the original sample, LR, \textbf{FAST ML}, and \textbf{synthpop} synthetic samples, respectively.}
  \label{t:reg_table}
  \resizebox{\textwidth}{!}{
  \begin{tabular}{llllllll}
    \hline
    & \textbf{Original}   & \textbf{LR $k$NN} & \textbf{LR $k$NN} & \textbf{LR K-Means} & \textbf{LR K-Means}  & \textbf{FAST ML}    & \textbf{Synthpop}    \\
    &   & \textbf{with MVN} & \textbf{with GC} & \textbf{with MVN} & \textbf{with GC}  &   &   \\
    \hline
    constant       & -0.4391 & -0.7364       & 0.0807       & \textbf{-0.4398} & -0.1604 & -0.2355 & 0.0104      \\
                   & (0.0276)   & (0.0281)         & (0.0309)        & (0.0280)   & (0.0276)   & (0.0282)   & (0.0263)    \\
    MedInc         & 0.5369  & 0.6087        & 0.5988       & \textbf{0.5347}  & 0.5036  & 0.5467  & 0.4566   \\
                   & (0.0041)   & (0.0042)         & (0.0041)        & (0.0041)   & (0.0038)   & (0.0042)   & (0.0034)    \\
    HouseAge       & 0.0165  & 0.0168        & 0.0173       & \textbf{0.0165}  & 0.0162  & 0.0164  & 0.0164   \\
                   & (0.0005)   & (0.0004)         & (0.0004)        & (0.0005)   & (0.0005)   & (0.0005)   & (0.0005)    \\
    AveRooms       & -0.2117 & -0.3124       & -0.2999      & \textbf{-0.2169} & -0.1465 & -0.1619 & -0.0851  \\
                   & (0.0060)   & (0.0064)         & (0.0065)        & (0.0060)   & (0.0052)   & (0.0041)   & (0.0042)    \\
    AveBedrms      & 0.9937  & 1.5311        & 1.2893       & \textbf{1.0218}  & 0.5438  & 0.5967  & 0.3262   \\
                   & (0.0295)   & (0.0329)         & (0.0307)        & (0.0298)   & (0.0227)   & (0.0185)   & (0.0174)    \\
    Population     & 0.0000  & 0.0000        & 0.0000       & 0.0000  & 0.0000  & 0.0000  & 0.0000   \\
                   & (0.0000)   & (0.0000)         & (0.0000)        & (0.0000)   & (0.0000)   & (0.0000)   & (0.0000)    \\
    AveOccup       & -0.0049 & -0.0127       & -0.2317      & -0.0059 & \textbf{-0.0052} & -0.0186 & -0.0320  \\
                   & (0.0005)   & (0.0011)         & (0.0057)        & (0.0006)   & (0.0007)   & (0.0007)   & (0.0022)    \\
    $R^2$      & 0.5397     & 0.6005           & 0.6328          & 0.5374     & \textbf{0.5346}     & 0.5583     & 0.5231      \\
    \hline
\end{tabular}

  }
  \vspace*{7pt}
  \scriptsize
  \begin{tablenotes}
      \item \textbf{Notes:} Bold values indicate the best approximations to the original sample coefficients for each variable. The abbreviations for the variable names are the following. {MedInc} is the median income in the (census) block, {HouseAge} is the median age of the houses in the block, {AveRooms} is the average number of rooms per household, {AveBedrms} is the average number of bedrooms per household, {Population} represents the number of people residing in the block, {AveOccup} denotes the average occupancy, representing the average number of individuals living in a single dwelling in the block, {R-squared} is the proportion of the variance in the dependent variable that's explained by the independent variables.
  \end{tablenotes}
\end{table}

The estimated coefficients shown in the first column are obtained from the original sample estimations. Subsequent columns display the coefficients estimated with the synthetic samples.\footnote{It is important to point out that the purpose of this regression analysis is not to estimate an accurate hedonic function, but to demonstrate that correlations in a multivariate analysis can be retained with synthetic samples.} First, results indicate that most coefficients from the synthetic samples are statistically different from those estimated from the original sample, with the exception of LR with the K-Means algorithm and MVN. Second, outliers could have a significant impact on the estimated coefficients. The disparity between the original sample coefficients and $k$NN-LR synthetic samples is partially driven by Theorem \ref{th:outlier} and the nature of the least squares estimations. This is evident by the observation that the synthetic samples created with the clustering algorithm yield a better approximation to the original sample estimations. Nevertheless, coefficients estimated with \textbf{FAST ML} and \textbf{synthpop} are also statistically different from the original sample, although they are not inherently biased like $k$NN-LR. Lastly, it is noteworthy that all the evaluated methods have room for improvement. The aim of this application is to demonstrate that LR approach can produce high-quality benchmark results with relatively simple configurations and implications of Theorem \ref{th:outlier} for a statistical analysis.

\end{document}